\documentclass{ws-procs9x6}

\def\beq{\begin{equation}}
\def\eeq{\end{equation}}

\begin{document}

\title{LORENTZ VIOLATION AND EXTENDED SUPERSYMMETRY}

\author{DON COLLADAY and PATRICK MCDONALD$^*$}

\address{Division of Natural Science, New College of Florida\\
Sarasota, FL 34234, USA\\
$^*$E-mail: mcdonald@ncf.edu}

%\author{A.N.\ AUTHOR}

%\address{Group, Laboratory, Street,\\
%City, State ZIP/Zone, Country\\
%E-mail: an\_author@laboratory.com}

\begin{abstract}
We construct a collection of Lorentz violating Yang-Mills theories
exhibiting supersymmetry. 
\end{abstract}

\bodymatter

\section{Introduction and background}
Symmetry has played a fundamental role in the construction of field
theories purporting to describe fundamental physics.  Nowhere is this
more true than in the construction of the Minimal Supersymmetric Standard
Model where symmetries mixing bosonic and fermionic states lead to
tightly constrained theories often exhibiting remarkable properties.
This is particularly true for ${\mathcal N}=4$ extended
supersymmetric  Yang-Mills theories \cite{BSS}, which are known to be
finite.

In this note we construct field theories which exhibit ${\mathcal N}=4$ 
extended supersymmetry and Lorentz violation.  To do so we
combine ideas of Berger and Kosteleck\'y \cite{BK} involving
supersymmetric scalar theories exhibiting Lorentz violation, and
well-known constructions of extended supersymmetric theories involving
dimensional reduction (nicely described in the work of Brink, Schwartz and
Scherk\cite{BSS}).  We begin by establishing notation in the context of
the standard construction.

Consider a gauge theory involving a single fermion $\lambda$ and
lagrangian 
\beq
{\mathcal L} = -\frac{1}{4} F^2 +\frac{i}{2}\bar{\lambda}\not \partial
\lambda ,
\eeq
where $F$ is the field strength, $F^{\mu\nu} =
[D^\mu,D^\nu]/ig$ and $D$ is the covariant derivative,
\beq
D^\mu = \partial^\mu +ig A^\mu.
\eeq
To simplify notation, we will first consider the abelian case.  To
implement supersymmetry we introduce a supercharge, $Q,$ satisfying 
\beq
[P_\mu,Q] = 0, \hspace{.5in} \{Q,\bar{Q}\} = 2\gamma^\mu P_\mu,
\eeq
where $\gamma^\mu$ are the standard Dirac matrices 
\beq
\{ \gamma^\mu,\gamma^\nu\} = 2 g^{\mu \nu},
\eeq
and the energy
momentum 4-vector $P_\mu$ generates spacetime translations.  The
construction of a supercharge is elegantly carried out in the context
of {\it superspace.}  More precisely, introduce four independent
anticommuting variables, $\theta,$ and consider the general vector
superfield 
\begin{eqnarray}
V(x,\theta)& = &C(x) + i\bar{\theta}\gamma^5 w(x)
-\frac{i}{2}\bar{\theta}\gamma^5\theta M(x)
-\frac{1}{2}\bar{\theta}\theta N(x)
+\frac{1}{2}\bar{\theta}\gamma^5\gamma^\mu \theta A_\mu  \nonumber\\
 &  & -i\bar{\theta}\gamma^5\theta \bar{\theta}[\lambda +\frac{i}{2} \not
  \partial w(x)] + \frac{1}{2}(\bar{\theta}\theta)^2(D(x)
-\frac{1}{2}\partial_\mu\partial^\mu C(x))
\end{eqnarray}
and the $Q$ operator
\beq
Q= -i \partial_{\bar{\theta}} -\gamma^\mu \theta \partial_\mu.
\eeq
Fixing an arbitrary spinor $\alpha,$ standard analysis\cite{S} of the operator
$\delta_Q V = -i\bar{\alpha}QV$ leads to the supersymmetry
transformations defining an ${\mathcal N}=1$ supersymmetric theory.

\section{${\mathcal N}=1$ supersymmetry}

Following Berger and Kosteleck\'y\cite{BK}, we introduce Lorentz
violation by defining a twisted derivative: 
\begin{equation}\label{perturbation}
\tilde{\partial}^\mu  =  \partial^\mu + k^{\mu \nu} \partial_\nu,
\end{equation}
where $k^{\mu \nu}$ is a symmetric, traceless, dimensionless tensor
parametrizing Lorentz violation.  To obtain a gauge invariant theory,
we also twist the underlying connection: 
\beq
\tilde{A}^\mu = A^\mu + k^{\mu \nu} A_\nu.
\eeq
These perturbations lead to a general vector superfield 
\begin{eqnarray}
\tilde{V}(x,\theta)& = &C(x) + i\bar{\theta}\gamma^5 w(x)
-\frac{i}{2}\bar{\theta}\gamma^5\theta M(x)
-\frac{1}{2}\bar{\theta}\theta N(x)
+\frac{1}{2}\bar{\theta}\gamma^5\gamma^\mu \theta \tilde{A}_\mu  \nonumber\\
 &  & -i\bar{\theta}\gamma^5\theta \bar{\theta}[\lambda +\frac{i}{2} \not
  \tilde{\partial} w(x)] + \frac{1}{2}(\bar{\theta}\theta)^2(D(x)
-\frac{1}{2}\tilde{\partial}_\mu\tilde{\partial}^\mu C(x))
\end{eqnarray}
and perturbed $Q$ operator
\beq
\tilde{Q}= -i \partial_{\bar{\theta}} -\gamma^\mu \theta \tilde{\partial}_\mu.
\eeq
Using Wess-Zumino gauge we obtain a lagrangian
\begin{equation}\label{lagrangian1}
\tilde{{\mathcal L}} = \frac{1}{4} \tilde{F}^2 +\frac{i}{2}
\bar{\lambda} \not \tilde{\partial} \lambda +\frac{1}{2} D^2,
\end{equation}
where $D$ is an auxiliary chiral field and $\tilde{F}$ is the twisted
field strength, $\tilde{F}^{\mu \nu} =\tilde{\partial}^\mu
\tilde{A}^\nu -\tilde{\partial}^\nu \tilde{A}^\mu.$  The twisted field
strength can be written in terms of the standard SME parameters: 
$\tilde{F}^2 = F^2 +k_F^{\mu\nu \alpha \beta}F_{\mu\nu}F_{\alpha
  \beta}$, where 
\begin{eqnarray}
k_{F}^{\mu \nu \alpha \beta} & = & 2(2k^{\alpha \mu} +(k^2)^{\alpha
  \mu})g^{\beta \nu} +4(k^{\mu \alpha} +(k^2)^{\alpha \mu})k^{\nu
  \beta} +(k^2)^{\alpha \mu} (k^2)^{\beta \nu}.
\quad
\end{eqnarray} 
Direct calculation confirms that the action is invariant under the
supersymmetry transformations 
\begin{eqnarray}
\delta \tilde{A}^\mu & = & -i\bar{\alpha} \gamma^\mu \lambda, \nonumber\\ 
\delta \lambda & = &
\frac{i}{2}\sigma^{\mu\nu}\tilde{F}_{\mu\nu}\alpha  - \gamma^5
D\alpha, \nonumber\\ 
\delta D & = & \bar{\alpha} \not \partial \gamma^5 \lambda.
\end{eqnarray}
This defines an ${\mathcal N}=1$ supersymmetric theory with Lorentz violation.  

\section{${\mathcal N}=4$ supersymmetry}

To build an ${\mathcal N}=4$ supersymmetric theory we work in $4+6$-dimensional
spacetime.  We represent the $32\times 32$ gamma matrices via
$\Gamma^\mu = \gamma^\mu \otimes I_8$ where $I_8$ is the $8\times 8$
identity matrix and $\mu = 0, 1, 2, 3, $ and 
\begin{eqnarray}
\Gamma^4 = \Gamma^{14} +\Gamma^{23}, \hspace{.25in} & \Gamma^6 = \Gamma^{34}
+\Gamma^{12}, &\hspace{.25in} i\Gamma^8 = \Gamma^{24} +\Gamma^{13}, \nonumber\\  
\Gamma^5 = \Gamma^{24} -\Gamma^{13},\hspace{.25in} & i\Gamma^7 = \Gamma^{14}
-\Gamma^{23}, &\hspace{.25in} i\Gamma^9 = \Gamma^{34} -\Gamma^{12}, 
\end{eqnarray}
where 
\beq
\Gamma^{ij} = \gamma_5 \otimes \left(\begin{array}{cc}
                                       0 & \rho^{ij} \\
                                       \rho_{ij} & 0 
                                      \end{array} \right)
\eeq
and the $4\times 4$ matrices $\rho$ are defined by 
\begin{eqnarray}
(\rho^{ij})_{kl} & = & \delta_{ik} \delta_{jl} -\delta_{jk}
  \delta_{il},\nonumber\\ 
(\rho_{ij})_{kl} & = & \frac{1}{2} \epsilon_{ijmn}(\rho^{mn})_{kl} =
  \epsilon_{ijkl}. 
\end{eqnarray}
We consider the lagrangian 
\beq
{\mathcal L} = -\frac{1}{4} \tilde{F}^2 + \frac{i}{2} \bar{\lambda}
\tilde{\not \partial} \lambda,
\eeq
where $\tilde{\not \partial} = \Gamma^\mu (\partial_\mu +
k_{\mu\nu}\partial^\nu) $ is the twisted derivative with $k_{\mu\nu}$
parametrizing $SO(1,9)$ violation and $\tilde{F}$ is the corresponding
perturbed field strength.  

Imposing both the Weyl and the Majorana condition and compactifying,
the fermion $\lambda$ satisfies  
\begin{eqnarray}
\lambda = \left(\begin{array}{cc}
                       L\chi \\
                       R\tilde{\chi} 
                 \end{array} \right) , &  & 
\end{eqnarray}
where $L$ and $R$ denote left and right projection operators,
respectively, the spinor $\chi$ transforms as a 4 of $SU(4)$ and the
(independent) spinor $\tilde{\chi}$ transforms as a $\bar{4}$ of $SU(4).$   

Choosing the Lorentz violating parameters with care leads to Lorentz
violating extended supersymmetric theories which are easy to describe.
For example, taking the $k_{\mu\nu}$ to vanish in the compactified
  directions leads to a lagrangian of the form 
\beq
{\mathcal L} = -\frac{1}{4} \tilde{F}^2 + i \bar{\chi}
\tilde{\not \partial} L\chi +\frac{1}{4} \tilde{\partial}_\mu
\phi_{ij}\tilde{\partial}^\mu \phi^{ij} ,
\eeq
where the complex scalar fields $\phi_{ij}$ transform as a 6 of
$SU(4)$ and are given by
\begin{eqnarray}
\phi_{i4}& = & \frac{1}{\sqrt{2}}(A_{i+3}+iA_{i+6}),\nonumber\\
\phi^{jk} & = &\frac{1}{2} \epsilon^{jklm}\phi_{lm} = (\phi_{jk})^*. 
\end{eqnarray}
The associated action is invariant under the supersymmetry
transformations 
\begin{eqnarray}
\delta \tilde{A}^\mu & = & -i(\bar{\alpha}_i \gamma^\mu L\chi^i
-\bar{\chi}_i \gamma^\mu L\alpha^i),   \nonumber\\
\delta \phi_{ij} & = & -i\sqrt{2}(\bar{\alpha}_j R\tilde{\chi}_i -
\bar{\alpha}_i R\tilde{\chi}_j
+\epsilon_{ijkl}\bar{\tilde{\alpha}}^k L \chi^l), \nonumber\\
\delta L\chi^i & = &
\frac{i}{2}\sigma^{\mu\nu}\tilde{F}_{\mu\nu}L\alpha^i -
\sqrt{2}\gamma^\mu \tilde{\partial}_\mu \phi^{ij} R\tilde{\alpha}_j, \nonumber\\
\delta R\tilde{\chi}_i & = &
\frac{i}{2}\sigma^{\mu\nu}\tilde{F}_{\mu\nu}R\tilde{\alpha}_i +
\sqrt{2}\gamma^\mu \tilde{\partial}_\mu \phi_{ij} L\alpha^j. 
\end{eqnarray}

Similarly, choosing the $k_{\mu \nu}$ parameters to vanish in the
spacetime directions $\mu = 0, 1, 2, 3,$ we obtain a Lagrangian of the
form 
\beq
{\mathcal L} = -\frac{1}{4} F^2 + i \bar{\chi}
\not \partial L\chi +\frac{1}{4} \partial_\mu
\tilde{\phi}_{ij}\partial^\mu \tilde{\phi}^{ij} ,
\eeq
where $\tilde{\phi}_{ij} = \phi_{ij} +\Lambda_{ijkl}\phi_{kl}$ with
the matrix $\Lambda_{ijkl}$ containing the effect of Lorentz violation
in the compactified directions.  The associated action is invariant
under the supersymmetry transformations

\begin{eqnarray}
\delta A^\mu & = & -i(\bar{\alpha}_i \gamma^\mu L\chi^i
-\bar{\chi}_i \gamma^\mu L\alpha^i),  \nonumber \\
\delta \tilde{\phi}_{ij} & = & -i\sqrt{2}(\bar{\alpha}_j R\tilde{\chi}_i -
\bar{\alpha}_i R\tilde{\chi}_j
+\epsilon_{ijkl}\bar{\tilde{\alpha}}^k L \chi^l), \nonumber \\
\delta L\chi^i & = &
\frac{i}{2}\sigma^{\mu\nu}F_{\mu\nu}L\alpha^i -
\sqrt{2}\gamma^\mu \partial_\mu \tilde{\phi}^{ij} R\tilde{\alpha}_j ,\nonumber \\
\delta R\tilde{\chi}_i & = &
\frac{i}{2}\sigma^{\mu\nu}F_{\mu\nu}R\tilde{\alpha}_i +
\sqrt{2}\gamma^\mu \partial_\mu \tilde{\phi}_{ij} L\alpha^j. 
\end{eqnarray}

Note that if the scalars $\tilde{\phi}_{ij}$ are identified with
physical scalars $\phi_{ij}$ we restore $SU(4)$ symmetry and remove
any Lorentz violating effects.  If, however, the $\phi_{ij}$ couple to
other sectors, Lorentz effects may reappear in these sectors.  

\section{Extensions and clarifications}

The above results warrant a number of additional comments:

\begin{itemize}
\item These constructions can be carried out in the nonabelian case
where they yield supersymmetric theories which exhibit
Lorentz violation\cite{cm}.  

\item The same techniques can be applied to obtain an ${\mathcal N}=2$ supersymmetric
theory with Lorentz violation.  The construction proceeds by working
in 4+2-dimensional spacetime and using dimensional reduction\cite{cm}.  

\item Because these constructions involve changing the structure of the
underlying superalgebra,\cite{cm,cm2} the no-go results of Nibbelink and Pospelov\cite{NP}
do not apply.

\end{itemize}

\end{document}